\documentclass[]{SciPost}

\hypersetup{
    colorlinks,
    linkcolor={red!50!black},
    citecolor={blue!50!black},
    urlcolor={blue!80!black}
}

\usepackage[bitstream-charter]{mathdesign}
\DeclareSymbolFont{usualmathcal}{OMS}{cmsy}{m}{n}
\DeclareSymbolFontAlphabet{\mathcal}{usualmathcal}

\DeclareMathOperator{\sign}{sign}

\begin{document}

\begin{center}{\Large \textbf{
High-dimensional manifold of solutions in neural networks: insights from statistical physics\\
}}\end{center}

\begin{center}
Enrico M. Malatesta\textsuperscript{1$\star$} 
\end{center}

\begin{center}
{\bf 1} Department of Computing Sciences, Bocconi University, 20136 Milano, Italy
\\
${}^\star$ {\small \sf enrico.malatesta@unibocconi.it}
\end{center}

\begin{center}
\today
\end{center}


\section*{Abstract}
{\bf
In these pedagogic notes I review the statistical mechanics approach to neural networks, focusing on the paradigmatic example of the perceptron architecture with binary an continuous weights, in the classification setting. I will review the Gardner's approach based on replica method and the derivation of the SAT/UNSAT transition in the storage setting. Then, I discuss some recent works that unveiled how the zero training error configurations are geometrically arranged, and how this arrangement changes as the size of the training set increases. I also illustrate how different regions of solution space can be explored analytically and how the landscape in the vicinity of a solution can be characterized. 
I give evidence how, in binary weight models, algorithmic hardness is a consequence of the disappearance of a clustered region of solutions that extends to very large distances.
Finally, I demonstrate how the study of linear mode connectivity between solutions can give insights into the average shape of the solution manifold.

}

\vspace{10pt}
\noindent\rule{\textwidth}{1pt}
\tableofcontents\thispagestyle{fancy}
\noindent\rule{\textwidth}{1pt}
\vspace{10pt}

\section{Introduction}

Suppose we are given a dataset $\mathcal{D} = \left\{ \boldsymbol{\xi}^\mu, y^\mu\right\}_{\mu = 1}^P$ composed by a set of $P$, $N$-dimensional ``patterns'' $\xi_i^\mu$, $i = 1, \dots, N$ and the corresponding label $y^\mu$. The patterns can represent whatever type of data, for example an image, text or audio. In the binary classification setting that we will consider, $y^\mu = \pm 1$; it will represent some particular property of the data that we want to be able to predict, e.g. if in the image there is a cat or a dog.  The goal is to learn a function $f: \mathbb{R}^N \to \pm 1$ that is able to associate to each input $\boldsymbol{\xi}\in \mathbb{R}^N$ the corresponding label. This function should be able to \emph{generalize}, i.e. to \emph{predict} the label corresponding to a pattern not in the training set.

In the following we will consider, in order to fit training set, the so-called \emph{perceptron} model, as it is the simplest, yet non-trivial, one-layer neural network that we can study using statistical mechanics tools. 
Given an input pattern $\boldsymbol{\xi}^\mu$ the perceptron predicts a label $\hat{y}^\mu$ 
\begin{equation}
	\hat{y}^\mu = \sign\left( \frac{1}{\sqrt{N}} \sum_{i=1}^{N} w_i \xi_i^\mu \right)
\end{equation}
where $w_i$ are the $N$ parameters that need to be adjusted in order to fit the training set. If the possible values of $\boldsymbol{w}$ lie on the vertices of the hypercube, i.e. $w_i = \pm 1$, $\forall i$, the model is called \emph{binary} perceptron. If instead the weights lie on the $N-$dimensional sphere of radius $\sqrt{N}$, i.e. $\sum_{i=1}^{N} w_i^2 = \boldsymbol{w} \cdot \boldsymbol{w} = N$ the model is called \emph{spherical} perceptron.

As usual in statistical mechanics, we consider the case of a \emph{synthetic} dataset, where the patterns are formed by random i.i.d. $N$-dimensional Gaussian patterns $\xi_i^\mu\sim \mathcal{N}(0, 1)$, $i = 1, \dots, N$. Depending on the choice of the label $y^\mu$, we can define two different scenarios:
\begin{itemize}
	\item in the so called \emph{teacher-student} scenario, $y^\mu$ is generated by a network, called ``teacher''. The simplest case corresponds to a perceptron architecture, i.e.
	\begin{equation}
		y^\mu = \text{sign}\left( \frac{1}{\sqrt{N}} \sum_{i=1}^N w_i^T \xi^\mu_i \right)
	\end{equation}
	with random i.i.d. Gaussian or binary weights: $w_i^T \sim \mathcal{N}(0, 1)$ or $w_i^T=\pm 1$ with equal probability. This model has been studied by Gardner and Derrida~\cite{Gardner_1989} and Gy\"orgyi in the binary case~\cite{Gyorgyi1990}.
	\item In real scenarios, sometimes data can be corrupted or the underline rule that generates the label for a given input is noisy. This makes the problem sometimes being \emph{unrealizable}, since, expecially for large datasets, no student exists that is able to learn the training set. The \emph{storage problem}, describes the case of extreme noise in the dataset: $y^\mu$ is chosen to be $\pm 1$ with equal probability. As a matter of fact this makes the label completely independent from the input data.
\end{itemize}


In the following we will focus on the storage problem setting. Even if this problem does not present the notion of generalization error, still, it is a scenario where we can try to answer very simple questions using the same statistical mechanics tools that can be also applied to teacher-student settings (see for example~\cite{engel-vandenbroek}). For example: what is the maximum value of the size of the training set $P$ that we are able to fit? It was discovered by the pioneering work by Gardner~\cite{gardner1987, gardner1988The} and Krauth and Mezard~\cite{krauth1989storage} that, in both binary and continuous settings, in the large $P$ and $N$ limit, with the ratio $\alpha \equiv P/N$ fixed, it exists a \emph{sharp} phase transition $\alpha_c$ separating a satisfiable (SAT) phase $\alpha < \alpha_c$ where solutions to the problem exists, to an unsatisfiable (UNSAT) phase where the set of solutions is empty. The goal of the next Section~\ref{sec::Gardner_computation} is to introduce the statistical mechanics tools through which we can answer such a question. The same tools, in turn, will provide valuable insights into how the solutions are arranged geometrically.


\section{Gardner's computation} \label{sec::Gardner_computation}

\subsection{Statistical mechanics representation}

Every good statistical mechanics computation starts by writing the partition function of the model under consideration. This is the goal of this section.

Fitting the training set means that we need to satisfy the following set of constraints
\begin{equation}
	\label{eq::zero_margin}
	\Delta^\mu \equiv \frac{y^\mu}{\sqrt{N}} \sum_i w_i \xi_i^\mu \ge 0\,, \qquad \mu = 1, \dots, P
\end{equation}
indeed if the \emph{stability} $\Delta^\mu$ of example $\mu$ in the training set is positive, it means that the predicted label is equal to the true one. In general we can try to ask to fit the training set in a way that is \emph{robust} with respect to noisy perturbation of the inputs; this can be done by requiring that the stability of each pattern is larger that a given threshold $\kappa \ge 0$ called \emph{margin}
\begin{equation}
	\label{eq::margin}
	\Delta^\mu \equiv \frac{y^\mu}{\sqrt{N}} \sum_i w_i \xi_i^\mu \ge \kappa\,, \qquad \mu = 1, \dots, P
\end{equation}
The total amount of noise that we can inject in $\boldsymbol{\xi}^\mu$ without changing the corresponding label $y^\mu$ depends on $\kappa$. Notice that if $\kappa \ge 0$ the solutions to~\eqref{eq::margin} are also solutions to~\eqref{eq::zero_margin}. 
Sometimes satisfying all the contraints~\eqref{eq::zero_margin} can be hard or even impossible, for example when the problem is not linearly separable. In that case one can relax the problem by counting as ``satisfied'' certain violated contraints: one way to do that is to impose a \emph{negative} margin $\kappa$. The corresponding model is usually called \emph{negative perceptron}. As we will see in the next subsection this changes dramatically the nature of the problem.

We are now ready to write down the partition function, as
\begin{equation}
	\label{eq::Z}
	Z_{\mathcal{D}} = \int d\mu(\boldsymbol{w}) \, \mathbb{X}_{\mathcal{D}}(\boldsymbol{w}; \kappa)
\end{equation}
where
\begin{equation}
	\label{eq::Gibbs}
	\mathbb{X}_{\mathcal{D}}(\boldsymbol{w}; \kappa) \equiv \prod_{\mu = 1}^{P}\Theta\left(\frac{y^\mu}{\sqrt{N}} \sum_i w_i \xi_i^\mu - \kappa\right)
\end{equation}
is an indicator function that selects a solution to the learning problem. Indeed $\Theta(x)$ is the Heaviside Theta function that gives 1 if $x>0$ and zero otherwise. $d \mu(\boldsymbol{w})$ is a measure over the weights that depends on the spherical/binary nature of the problem under consideration
\begin{equation}
	\label{eq::measure_weights}
	d\mu(\boldsymbol{w}) = \left\{ 
	\begin{array}{ll}
		\prod_{i = 1}^{N} dw_i \, \delta\left( \boldsymbol{w} \cdot \boldsymbol{w} 
		- N\right), & \qquad\text{\emph{spherical} case} \\
		\prod_{i = 1}^{N} dw_i \, \prod_i\left[\delta\left( w_i - 1\right) + \delta\left( w_i + 1\right)  \right], & \qquad\text{\emph{binary} case}
	\end{array}
	\right.
\end{equation}
The partition function is called also \emph{Gardner} volume~\cite{gardner1988The} since it measures the total volume (or number in the binary case) of networks satisfying all the constraints imposed by the training set. 
Notice that sending $\xi_i^\mu \to y^\mu \xi_i^\mu$ one does not change the probability measure over the patterns $\boldsymbol{\xi}^\mu$, therefore we can simply set $y^\mu = 1$ for all $\mu = 1, \dots, P$ without loss of generality.

\subsection{Simple geometric properties of the solution space}
\begin{figure}[h]
	\begin{centering}
		\includegraphics[width=0.3\columnwidth]{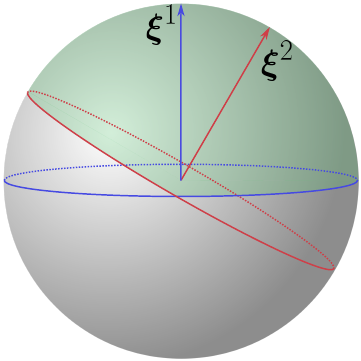}
		\hfill 
		\includegraphics[width=0.3\columnwidth]{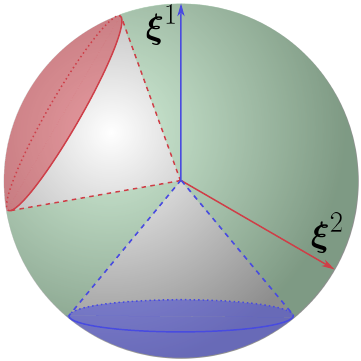}
		\hfill
		\includegraphics[width=0.3\columnwidth]{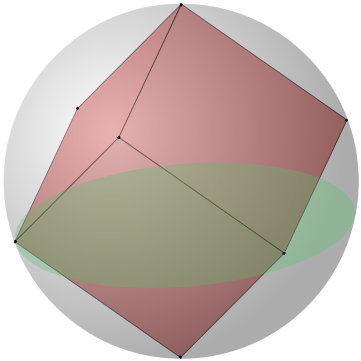}
	\end{centering}
	\caption{{\bf Left and middle panels}: space of solutions (green shaded area) in the spherical perceptron with $\kappa = 0$ (left panel) and $\kappa < 0$ (middle). The green shaded area is obtained by taking the intersection of two spherical caps corresponding to patterns $\boldsymbol{\xi}^1$ and $\boldsymbol{\xi}^2$. {\bf Right panel}: binary weight case. The cube (red) is inscribed in the sphere. The green shaded area represents the (convex) space of solutions of the corresponding spherical problem. In this simple example only two vertices of the cube are inside the green region.}
	\label{Fig::patterns_on_sphere} 
\end{figure}

We want here to give a simple geometrical interpretation of the space of solutions corresponding to imposing constraints of equation~\eqref{eq::margin}. By those simple arguments, we will be able to unravel the convexity/non-convexity properties of the space of solutions. 

Let's start from the spherical case. For simplicity consider also $\kappa = 0$. Initially, when no pattern has been presented, the whole volume of the $N$-dimensional sphere is a solution. When we present the first pattern $\boldsymbol{\xi}^1$ uniformly generated on the sphere, the allowed solutions lie on the half-sphere with a positive dot product with $\boldsymbol{\xi}^1$. Of course, the same would happen if we had presented any other pattern $\boldsymbol{\xi}^\mu$ of the training set. The intersection of those half-spheres form the space of solutions. Since the intersection of half spheres is a convex set, it turns out that the manifold of solutions is always convex, see left panel of Fig.~\ref{Fig::patterns_on_sphere} for an example. Notice that this is also true if one is interested in looking to the subset of solutions having $\kappa>0$. If one keeps adding constraints one can obtain an empty set. The minimal density of constraints for which this happens is the SAT/UNSAT transition.  

The middle panel of Fig.~\ref{Fig::patterns_on_sphere}, refers to the $\kappa <0$ case, i.e. the spherical negative perceptron~\cite{franz2017}. In this problem the space of solutions can be obtained by \emph{removing} from the sphere a convex spherical cap, one for each pattern (blue and red regions). As a result the manifold of solutions is non-convex and, if one adds too many constraints the space of solutions can become also \emph{disconnected}, before the SAT/UNSAT transition.  

In the right panel of Fig.~\ref{Fig::patterns_on_sphere} we show the binary case. Since $w_i = \pm 1$, the hypercube (in red) is inscribed in the sphere in $N$-dimension, i.e. the vertices of the hypercube are contained in the space of solutions of the corresponding spherical problem (green region). As can be readily seen from the example in the figure, the binary weight case is a non-convex problem (even for $\kappa \ge 0$), since in order to go from one solution to another, it can happen that one has to pass through a set of vertices not in the solution space.

\subsection{The typical Gardner volume}
The Gardner volume $Z_{\mathcal{D}}$ introduced in equation~\ref{eq::Z} is random variable since it explicitly depends on the dataset. The goal of statistical mechanics is to characterize the \emph{typical}, i.e. the most probable value of this random quantity. A first guess would be to compute the averaged volume $\langle Z_{\mathcal{D}} \rangle_{\mathcal{D}}$. However, since~\eqref{eq::Z} involves a product of many random contributions, the probability distribution for large $N$ tends to be log-normal, and the most probable value of $Z_{\mathcal{D}}$ and its average 
do not coincide\footnote{In the spin glass literature one says that $Z_{\mathcal{D}}$ is \emph{not} a \emph{self-averaging} quantity for large $N$.}. 
On the other hand, the log of the product of independent random variables is equivalent to a large sum of independent terms, that, because of the central limit theorem, is Gaussian distributed; in that case we expect that the most probable value coincides with the average. Therefore we expect that for large $N$
\begin{equation}
	\label{eq::Z_and_phi}
	Z_{\mathcal{D}} \sim e^{N \phi}
\end{equation}
where
\begin{equation}
	\phi = \lim\limits_{N \to \infty} \frac{1}{N} \langle \ln Z_{\mathcal{D}} \rangle_{\mathcal{D}} \,.
\end{equation}
is the averaged log-volume. Since we are at zero training error $\phi$ coincides with the \emph{entropy} of solutions. 
Performing the average over the log is usually called as \emph{quenched} average in spin glass theory, to distinguish from the log of the average, which is instead called \emph{annealed}. Annealed averages are much easier than quenched ones; even if they do not give information to the typical value of a random variable, they can still be useful, since they can give an upper bound to the quenched entropy. Indeed due to Jensen's inequality
\begin{equation}
	\phi \le \phi_{ann} = \lim\limits_{N\to \infty} \frac{1}{N} \ln \langle Z_{\mathcal{D}}\rangle_{\mathcal{D}} \,.
\end{equation}

\subsection{Replica Method} \label{sec::Replica_Method}
In order to compute the average of the log we use the \emph{replica trick}
\begin{equation}
	\langle \ln Z_{\mathcal{D}} \rangle_{\mathcal{D}} = \lim\limits_{n \to 0} \frac{\langle Z_{\mathcal{D}}^n\rangle_{\mathcal{D}} - 1}{n} = \lim\limits_{n \to 0} \frac{1}{n}\ln \langle Z_{\mathcal{D}}^n\rangle_{\mathcal{D}} \,.
\end{equation}
The \emph{replica method} consists in performing the average over the disorder of $Z^n_{\mathcal{D}}$ considering $n$ integer (a much easier task with respect to averaging the log of $Z_{\mathcal{D}}$), and then performing an analytic continuation of the result to $n\to 0$~\cite{mezard1987spin}. It has been used firstly in spin glasses models such as the Sherrington-Kirkpatrick model~\cite{Sherrington1975} and then applied by E. Gardner to neural networks~\cite{gardner1988The,gardner1988optimal}. Replicating the partition function and introducing an auxiliary variable $v_\mu^a \equiv \frac{1}{\sqrt{N}} \sum_i w_i^a \xi_i^\mu$ using a Dirac delta function, we have
\begin{equation}
	\begin{split}Z_{\mathcal{D}}^{n} & =\int\prod_{a=1}^{n}d\mu(w^{a})\,\prod_{\mu a}\Theta\left(\frac{1}{\sqrt{N}}\sum_i w_{i}^{a}\xi_{i}^{\mu}-\kappa\right)\\
		& =\int\prod_{a\mu}\frac{dv_{a}^{\mu}d\hat{v}_{a}^{\mu}}{2\pi}\,\prod_{\mu a}\Theta\left(v_{a}^{\mu}-\kappa\right)e^{iv_{a}^{\mu}\hat{v}_{a}^{\mu}}
		 \int\prod_{a=1}^{n}d\mu(\boldsymbol{w}^{a})\,e^{-i\sum_{\mu,a}\hat{v}_{a}^{\mu}\frac{1}{\sqrt{N}}\sum_{i}w_{i}^{a}\xi_{i}^{\mu}}
	\end{split}
\end{equation}
where we have used the integral representation of the Dirac delta function
\begin{equation}
	\delta(v) = \int \frac{d \hat v}{2\pi} e^{i v \hat v}\,.
\end{equation} 
Now we can perform the average over the patterns in the limit of
large $N$ obtaining 
\begin{equation}
	\begin{split}
		\prod_{\mu i}\left\langle e^{-i\frac{\xi_{i}^{\mu}}{\sqrt{N}}\sum_{a}w_{i}^{a}\hat{v}_{a}^{\mu}}\right\rangle _{\boldsymbol{\xi}^{\mu}_i} & =\prod_{\mu i}e^{ - \frac{1}{2N}\left( \sum_a w_i^a \hat v_\mu^a\right)^2} \\
		&=\prod_{\mu}e^{-\sum_{a<b}\hat{v}_{a}^{\mu}\hat{v}_{\mu}^{b}\left(\frac{1}{N}\sum_{i}w_{i}^{a}w_{i}^{b}\right)-\frac{1}{2}\sum_{a}\left(\hat{v}_{a}^{\mu}\right)^{2}} \,.
	\end{split}
\end{equation}
Next we can enforce the definition of the $n \times n$ matrix of order parameters 
\begin{equation}
	\label{eq::qab}
	q_{ab}\equiv\frac{1}{N}\sum_{i=1}^{N}w_{i}^{a}w_{i}^{b}
\end{equation}
by using a delta function and its integral representation. $q_{ab}$ represents the typical overlap between two replicas $a$ and $b$, sampled from the Gibbs measure corresponding to~\eqref{eq::Z} and having the same realization of the training set. Due to the binary and spherical normalization of the weights this quantity is bounded $-1 \le q_{ab} \le 1$. We have
\begin{equation}
	\begin{split}
		\langle Z_{\mathcal{D}}^n \rangle_{\mathcal{D}} &=\int \prod_{a<b} \frac{dq_{ab} d\hat q_{ab}}{2\pi/N}  e^{-N\sum_{a<b} q_{ab} \hat q_{ab}}  \int\prod_{a}d\mu(\boldsymbol{w}^{a})\,e^{\sum_{a<b} \hat q_{ab}\sum_i w_i^a w_i^b}\\
		&\times 
		\int\prod_{a\mu}\frac{dv_{a}^{\mu}d\hat{v}_{a}^{\mu}}{2\pi}\,\prod_{\mu a}\Theta\left(v_{a}^{\mu}-\kappa\right)e^{i \sum_{a\mu } v_{a}^{\mu}\hat{v}_{a}^{\mu} -\frac{1}{2}\sum_{a \mu} \left(\hat{v}_{a}^{\mu}\right)^{2} -\sum_{a<b, \mu} q_{ab} \hat{v}_{a}^{\mu}\hat{v}_{\mu}^{b}}	\,.
	\end{split}
\end{equation}
The next step of every replica computation is to notice that the terms depending on the patterns $\mu = 1, \dots, P$ and on the index of the weights $i = 1, \dots, N$ have been decoupled (initially they were not), at the price of coupling the replicas (that initially were uncopled). So far the computation was the same independently on the nature of the weights. From now on, however the computation is different, since we need to explicit the form of the measure $d \mu(\boldsymbol{w})$ in order to factorize the expression over the index $i$. In the next paragraph we therefore focus on the binary case first, moving then to the spherical case. 

\paragraph{Binary case}
Setting $P = \alpha N$, in the binary weight case we reach the following integral representation of the averaged replicated partition function
\begin{equation}
	\left\langle Z_{\mathcal{D}}^{n}\right\rangle _{\mathcal{D}} \propto\int\prod_{a<b}\frac{dq_{ab}d\hat{q}_{ab}}{2\pi}\,e^{NS(q,\,\hat{q})}
\end{equation}
where we have defined 
\begin{subequations} 
	\begin{align}
		S^{\text{bin}}(q,\hat{q}) & = G^{\text{bin}}_{S}(q, \hat{q})+\alpha G_{E}\left(q\right)\\
		G^{\text{bin}}_{S} & = -\frac{1}{2}\sum_{a\ne b}q_{ab}\hat{q}_{ab} + \ln\sum_{\left\{ w^{a}\right\} }e^{\frac{1}{2}\sum_{a\ne b}\hat{q}_{ab}w^{a}w^{b}}\label{Gs}\\
		G_{E} & =\ln \int\prod_{a}\frac{dv_{a} d\hat{v}_{a}}{2\pi}\,\prod_{a}\Theta\left(v_{a}-\kappa\right)e^{i \sum_a v_{a}\hat{v}_{a} - \frac{1}{2}\sum_{a b} q_{ab} \hat{v}_{a}\hat{v}_{b} } \,.
		\label{eq::Ge}
	\end{align}
\end{subequations}
$G_S$ is the so called ``entropic'' term, since it represents the logarithm of the volume at $\alpha = 0$, where there are no constraints induced by the training set. $G_E$ is instead called the ``energetic'' term and it represents the logarithm of the fraction of solutions. In the energetic term we have also used the fact that, because of equation~\eqref{eq::qab}, $q_{aa} = 1$.

\paragraph{Spherical case}
In the spherical case, one needs to do a little more work in order to decouple the $i$ index in the spherical constraints of equation~\eqref{eq::measure_weights}:
\begin{equation}
	\begin{split}
	\int\prod_{a}d\mu(w^{a}) \, e^{\sum_{a<b} \hat q_{ab}\sum_i w_i^a w_i^b} &\propto \int \prod_a \frac{d \hat q_{aa}}{2\pi}  \int \prod_{ai} dw_i^a \, e^{-\frac{N}{2} \sum_a \hat q_{aa} + \frac{1}{2}\sum_{ab} \hat q_{ab}\sum_i w_i^a w_i^b} \\
	&= \int \prod_a \frac{d \hat q_{aa}}{2\pi} e^{-\frac{N}{2} \sum_a \hat q_{aa} } \left[ \int \prod_a dw^a \, e^{\frac{1}{2}\sum_{ab} \hat q_{ab} w^a w^b} \right]^N
	\end{split}
\end{equation}
 Therefore the replicated averaged partition in the spherical case is equal to
 \begin{equation}
 	\left\langle Z_{\mathcal{D}}^{n}\right\rangle _{\mathcal{D}} \propto\int\prod_{a<b}\frac{dq_{ab}d\hat{q}_{ab}}{2\pi} \int \frac{d \hat q_{aa}}{2\pi}\,e^{NS(q,\,\hat{q})}
 \end{equation}
 where we have defined 
 \begin{subequations} 
 	\begin{align}
 		S^{\text{sph}}(q,\hat{q}) & = G^{\text{sph}}_{S}(\hat{q})+\alpha G_{E}\left(q\right)\\
 		G^{\text{sph}}_{S} & = -\frac{1}{2}\sum_{a b}q_{ab}\hat{q}_{ab}+\ln \int \prod_a dw^a \, e^{\frac{1}{2}\sum_{a b}\hat{q}_{ab}w^{a}w^{b}} \label{Gs_sph} \\
 		&= -\frac{1}{2}\sum_{a b}q_{ab}\hat{q}_{ab} - \frac{1}{2}\ln \det (-\hat q)  \nonumber
 	\end{align}
 \end{subequations}
 and we have used again the definition $q_{aa} = 1$. Notice how only the entropic term changes with respect to the binary case, whereas the energetic term is the same defined in~\eqref{eq::Ge}.
 
\paragraph{Saddle points}
In either case, being the model binary or spherical, the corresponding expressions can be evaluated by using a saddle point approximation, since we are interested in a regime where $N$ is large. The saddle point have to be found by finding two $n\times n$ matrices $q_{ab}$ and $\hat{q}_{ab}$ that maximize the action $S$. This gives access to the entropy $\phi$ of solutions
\begin{equation}
	\phi= \lim\limits_{N \to \infty} \frac{1}{N} \langle \ln Z_{\mathcal{D}} \rangle_{\mathcal{D}} = \lim\limits _{n\to0}\frac{1}{n}\max\limits _{\left\{ q,\hat{q}\right\} }S(q,\hat{q})\,.
\end{equation}

\subsection{Replica-Symmetric ansatz} \label{sec::RS}
Finding the solution to the maximization procedure is not a trivial task. Therefore one proceeds by formulating a simple parameterization or \emph{ansatz} on the structure of the saddle points. The simplest ansatz one can formulate is the Replica-Symmetric (RS) one. Considering the binary case, this reads
\begin{subequations}
	\begin{align}
		q_{ab} &=\delta_{ab}+q(1-\delta_{ab})\,,\\
		\hat{q}_{ab} &=\hat{q}(1-\delta_{ab})		
	\end{align}
\end{subequations}
In the spherical case, instead one has ($q_{ab}$ is the same as in binary weights)
\begin{equation}
	\label{eq::entropy}
	\hat{q}_{ab} = - \hat Q \delta_{ab} + \hat{q}(1-\delta_{ab})	
\end{equation}
The entropy in both cases can be written as
\begin{equation}
	\phi=\mathcal{G}_{S}+\alpha\mathcal{G}_{E}
\end{equation}
where we remind that the entropic term depends on the binary or spherical nature of the weights and are
\begin{subequations} 
	\begin{align}
		\mathcal{G}^{\text{bin}}_{S} & \equiv \lim\limits _{n\to0}\frac{G^{\text{bin}}_{S}}{n}=-\frac{\hat{q}}{2}(1-q)+\int Dx\ln2\cosh(\sqrt{\hat{q}}x)\,,\\
		\mathcal{G}^{\text{sph}}_{S} & \equiv \lim\limits _{n\to0}\frac{G^{\text{sph}}_{S}}{n}=\frac{1}{2} \hat Q + \frac{q \hat q}{2} + \frac{1}{2} \ln \frac{2\pi}{\hat Q + \hat q } + \frac{1}{2} \frac{\hat q}{\hat Q + \hat q }\,.
	\end{align}
\end{subequations}
The energetic term is instead common to both models
\begin{equation}
	\label{eq::GE_RS}
	\mathcal{G}_{E} \equiv\lim\limits _{n\to0}\frac{G_{E}}{n}=\int Dx\ln H\left(\frac{\kappa+\sqrt{q}x}{\sqrt{1-q}}\right) \,.
\end{equation}
In the previous equations we denoted by $Dx \equiv \frac{dx}{\sqrt{2\pi}} e^{-x^2/2}$ the standard normal measure and we have defined the function
\begin{equation}
	H(x) \equiv \int_x^\infty Dy = \frac{1}{2}\mathrm{Erfc}\left(\frac{x}{\sqrt{2}}\right)	
\end{equation}
The value of the order parameters $\hat q$, $\hat{Q}$ and $q$ for the spherical case and $\hat q$, $q$ for the binary case can be obtained by differentiating the entropy and equating it to zero.

\paragraph{Saddle point equations: spherical perceptron} 
The saddle point equations for the spherical case read
\begin{subequations}
	\label{eq::SP_sph}
	\begin{align}
		q &= \frac{\hat q}{(\hat Q + \hat q)^2} \label{eq::sp_sph_qh}\\
		1 &= \frac{\hat Q + 2\hat q}{(\hat Q + \hat q)^2} \label{eq::sp_sph_Qh}\\
		\hat{q} &= - 2\alpha \frac{\partial \mathcal{G}_E}{\partial q} = \frac{\alpha}{1-q}  \int Dx \, \left[GH\left(\frac{\kappa + \sqrt{q} x}{\sqrt{1-q}}\right) \right]^2 \label{eq::sp_sph_q}
	\end{align}
\end{subequations}
where we have defined the function $GH(x) \equiv \frac{G(x)}{H(x)}$, $G(x)$ being a standard normal distribution. The saddle point equations can be simplified, by explicitly expressing $\hat q$ and $\hat Q$ in terms of $q$. Indeed equations~\eqref{eq::sp_sph_qh}~\eqref{eq::sp_sph_Qh} are simple algebraic expression for $\hat q$ and $\hat{Q}$ in terms of $q$; they can be explicitly inverted as $\hat q = \frac{q}{(1-q)^2}$, $\hat Q = (1-2q) / (1-q)^2$. Inserting the expression of $\hat q$ inside~\eqref{eq::sp_sph_q} we get an equation for $q$ only
\begin{equation}
	\label{eq::SP_sph_q_only}
	q= \alpha \, (1-q)  \int Dx \, \left[GH\left(\frac{\kappa + \sqrt{q} x}{\sqrt{1-q}}\right) \right]^2 \,,
\end{equation}

\paragraph{Saddle point equations: binary perceptron} 

In the binary case only the equation involving derivatives of the entropic term changes. The saddle point equations are therefore
\begin{subequations}
	\label{eq::SP_bin}
	\begin{align}
		q &= \int Dx \, \tanh^2\left(\sqrt{\hat q} x\right) \\
		\hat{q} &= - 2\alpha \frac{\partial \mathcal{G}_E}{\partial q} = \frac{\alpha}{1-q}  \int Dx \, \left[GH\left(\frac{\kappa + \sqrt{q} x}{\sqrt{1-q}}\right) \right]^2 
	\end{align}
\end{subequations}
The saddle point equations (equation~\eqref{eq::SP_sph_q_only} for the spherical and~\eqref{eq::SP_bin} for the binary case), can be easily solved numerically by simple recursion with $\alpha$ and $\kappa$ being two external parameters. It is important to notice that the value of the order parameter $q$ has physical meaning. Indeed one can show that $q$ is the \emph{typical} (i.e. the most probable) overlap between two solutions $\boldsymbol{w}^1$ and $\boldsymbol{w}^2$ extracted from the Gibbs measure~\eqref{eq::Gibbs}, i.e.
\begin{equation}
	\label{eq::overlap}
	q = \left\langle \frac{\int d \mu(\boldsymbol{w}^1) \,  d \mu(\boldsymbol{w}^2) \, \left( \frac{1}{N} \sum_i w_i^1 w_i^2 \right) \mathbb{X}_{\mathcal{D}}(\boldsymbol{w}^1; \kappa) \mathbb{X}_{\mathcal{D}}(\boldsymbol{w}^2; \kappa)}{Z_{\mathcal{D}}^2} \right\rangle_{\mathcal{D}}
\end{equation}
Therefore solving the saddle point equations gives access to interesting geometrical information: it suggests how distant the solutions extracted from the Gibbs measure~\eqref{eq::Gibbs} are from each other. The distance between solutions can be obtained from the overlap using the relation
\begin{equation}
	\label{eq::distance_overlap_relation}
	d = \frac{1-q}{2} \in [0,1]
\end{equation}
\begin{figure}[t]
	\centering
	\includegraphics[width=0.49\columnwidth]{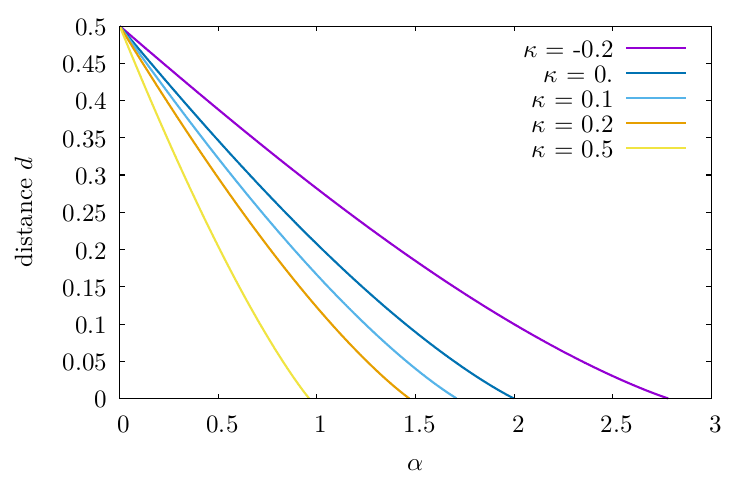}
	\includegraphics[width=0.49\columnwidth]{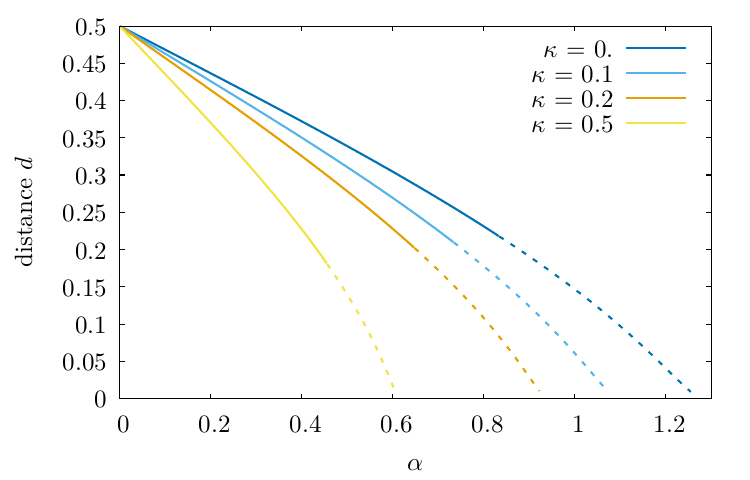}
	\caption{Distance between solutions sampled with a given margin $\kappa$ as a function of the constrained density $\alpha$ for the spherical case (\textbf{left panel}) and the binary perceptron (\textbf{right panel}). 
		In the binary case, the lines change from solid to dashed when the entropy of solutions becomes negative.
	}
	\label{Fig::distance} 
\end{figure}
In the binary setting this definition coincides with the \emph{Hamming distance} between $\boldsymbol{w}^1$ and $\boldsymbol{w}^2$, since in that case $d$ is equal to the fraction of indexes $i$ at which the corresponding $w_i^1$ and $w_i^2$ are different.
The distance between solutions extracted from the Gibbs measure with a given margin $\kappa$ is shown as a function of $\alpha$ in Fig.~\ref{Fig::distance} for the spherical (left panel) and for the binary (right panel) case. In both cases one can clearly see that the distance is monotonically decreasing with $\alpha$ it also exists a value of $\alpha$ for which the distance goes to zero. 
In addition is interesting to notice that fixing the value of $\alpha$ to a certain value, the distance decreases as the margin is increased, meaning that more robust solutions are located in a smaller region of the solution space~\cite{baldassi2021unveiling}.

\subsection{SAT/UNSAT transition}
\begin{figure}[t]
	\begin{centering}
		\includegraphics[width=0.493\columnwidth]{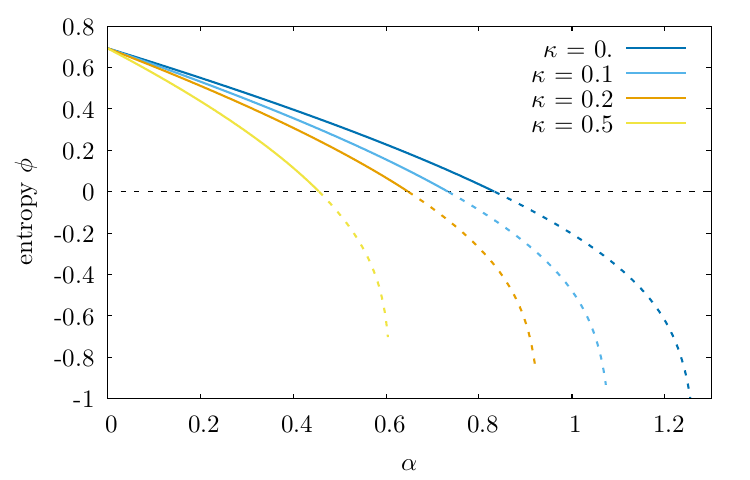}
		\includegraphics[width=0.493\columnwidth]{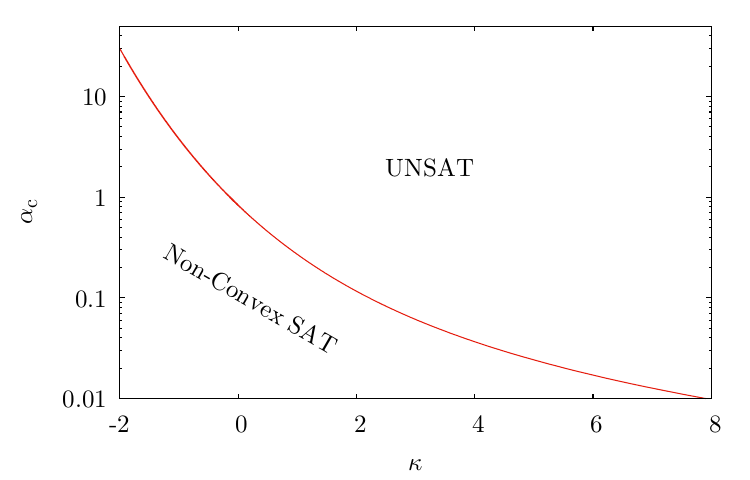}
	\end{centering}
	\caption{\textbf{Binary perceptron}. {\bf{Left}}: RS entropy as a function of $\alpha$ for several values of the margin $\kappa$. Dashed lines show the nonphysical parts of the curves, where entropy is negative. The value of $\alpha$ where the entropy vanishes corresponds to the SAT/UNSAT transition $\alpha_c(\kappa)$. {\bf{Right}}: SAT/UNSAT transition $\alpha_c(\kappa)$ obtained using the zero entropy condition~\eqref{eq::zero_entropy_condition}.}
	\label{Fig::binary_entropy_and_alphac} 
\end{figure}

\begin{figure}[t]
	\begin{centering}
		\includegraphics[width=0.493\columnwidth]{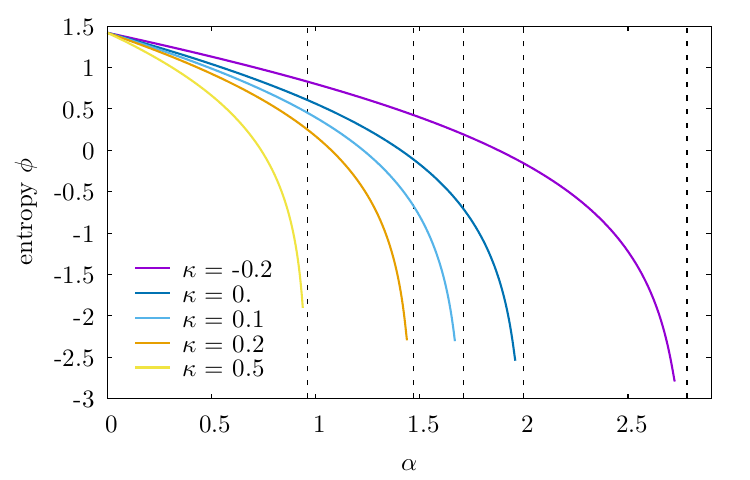}
		\includegraphics[width=0.493\columnwidth]{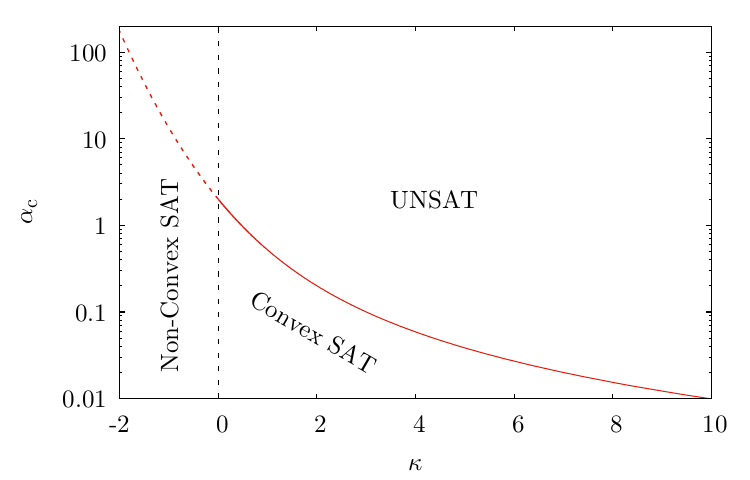}
	\end{centering}
	\caption{\textbf{Spherical perceptron}. {\bf{Left}}: RS entropy as a function of $\alpha$ for several values of the margin $\kappa$. 
		The point in $\alpha$ where the entropy goes to $-\infty$ (indicated by the dashed vertical lines) corresponds to the SAT/UNSAT transition $\alpha_c(\kappa)$. {\bf{Right}}: RS SAT/UNSAT transition $\alpha_c(\kappa)$ as given by~\eqref{eq::alphac_spherical}. For $\kappa<0$ the line is dashed, to remind that the RS prediction is only an upper bound to the true one, since the model becomes non-convex.}
	\label{Fig::spherical_entropy_and_alphac} 
\end{figure}

Once the saddle point equations are solved numerically, we can compute the value of the entropy (i.e. the total number/volume) of solutions using~\eqref{eq::entropy} for a given $\alpha$ and margin $\kappa$. The entropy is plotted in the left panel of Fig.~\ref{Fig::binary_entropy_and_alphac} and Fig.~\ref{Fig::spherical_entropy_and_alphac} for the binary and spherical cases respectively.

It is interesting to notice that in both models, for a fixed value of $\kappa$, there is a critical value of $\alpha$ such that the typical distance between solutions goes to zero, i.e. the solution space shrinks to a point as we approach it. The corresponding entropy diverges to $-\infty$ at this value of $\alpha$. This defines what we have called SAT/UNSAT transition $\alpha_c(\kappa)$ in the spherical case. This \emph{cannot} be the true value of $\alpha_c(\kappa)$ in the binary case: in binary models the entropy cannot be negative, since we are \emph{counting} solutions (not measuring volumes as in the spherical counterpart). This means that the analytical results obtained are wrong whenever $\phi < 0$; for this reason in Fig~\ref{Fig::distance} and~\ref{Fig::spherical_entropy_and_alphac} the unphysical parts of the curves are dashed. Where is the computation wrong in the binary case? As shown by~\cite{krauth1989storage}, it is the RS ansatz that, although stable, is wrong; at variance with the spherical case for $\kappa \ge 0$, in the binary case the solution space is non-convex, so the solution space can be disconnected; in those cases it is well known that the RS ansatz might fail. As shown by Krauth and M\'ezard~\cite{krauth1989storage} (see also~\cite{engel-vandenbroek} for a nice ``discussion''), by using a \emph{one-step replica symmetry breaking} (1RSB) ansatz~\cite{Parisi1980}, in order to compute the SAT/UNSAT transition we should compute the value of $\alpha$ for which the \emph{RS entropy} vanishes
\begin{equation}
	\label{eq::zero_entropy_condition}
	\phi^{\text{RS}}(\alpha_c) = 0\,.
\end{equation}
This is known as the \emph{zero entropy condition}; at this value of $\alpha_c$, the distance between solutions does not go to 0, for $\kappa = 0$ for example $d\simeq0.218$. In the right panel of Fig.~\ref{Fig::binary_entropy_and_alphac} we plot $\alpha_c$ as a function of $\kappa$. In particular for $\kappa = 0$, the value $\alpha_c = 0.833\dots$ can be obtained numerically. This value is still not rigorously proved, although in recent years~\cite{ding2019capacity}, proved that the value obtained by the replica formula is lower bound with positive probability. 

In the spherical perceptron $\kappa \ge 0$ the space of solution is convex and the RS ansatz gives correct results. The critical capacity can be therefore obtained by sending $q \to 1$. In order to get an explicit expression that can be evaluated numerically, one therefore has to do the limit explicitly, so a a little bit of work has still to be done.

The entropy is, in both cases, a monotonically decreasing function of $\alpha$. This should be expected: when we increase the number of contraints, we should expect that the solution space shrinks. Moreover for a fixed value of $\alpha$ the entropy is a monotonically decreasing function of the margin: this means that solutions with larger margin are \emph{exponentially} fewer in $N$ (remind~\eqref{eq::Z_and_phi}).

\paragraph{$q\to 1$ limit in the spherical perceptron}

In order to perform the limit analytically~\cite{gardner1987, baldassi2023typical} it is convenient to use the following change of variables in the entropy
\begin{equation}
	q = 1 - \delta q
\end{equation}
and then send $\delta q \to 0$. We now insert this into the RS energetic term~\eqref{eq::GE_RS}; using the fact that $\ln H(x) \simeq - \frac{1}{2} \ln(2\pi) - \ln x - \frac{x^2}{2}$ as $x \to \infty$, retaining only the diverging terms we get
\begin{equation}
	\begin{split}
		&\int Dx \ln H\left(\frac{\kappa + \sqrt{q} x}{\sqrt{1-q}}\right) \simeq \int_{-\kappa}^{+\infty} Dx \left[ \frac{1}{2} \ln \delta q - \frac{\left( \kappa + x \right)^2}{2 \delta q} \right]=\frac{1}{2} \ln (\delta q) \, H\left( -\kappa \right) - \frac{B(\kappa)}{2 \delta q}
	\end{split}
\end{equation}
and
\begin{equation}
	B(\kappa) = \int_{-\kappa}^{\infty} Dz_0 \, (\kappa + z_0)^2 = \kappa G\left( \kappa \right) + \left( \kappa^2 + 1 \right) H\left( -\kappa \right) \,.
\end{equation}
The free entropy is therefore
\begin{equation}
	\label{eq::phi}
	\phi =  \frac{1}{2 \delta q} + \frac{1}{2} \ln \delta q + \frac{\alpha}{2} \left( \ln (\delta q) \, H\left( -\kappa \right) - \frac{B(\kappa)}{\delta q} \right)
\end{equation}
The derivative with respect to $\delta q$ gives the saddle point condition that $\delta q$ itself must satisfy
\begin{equation}
	2 \frac{\partial \phi}{\partial \delta q} = \frac{1}{\delta q} - \frac{1}{\delta q^2} + \alpha \left( \frac{H\left( -\kappa \right)}{\delta q} + \frac{B(\kappa)}{\delta q^2}\right) = 0 \,.
\end{equation}
When we send $\delta q \to 0$, for a fixed value of $\kappa$ we can impose that we are near the critical capacity $\alpha = \alpha_c - \delta \alpha$ and $\delta q = C \delta \alpha$. We get
\begin{equation}
	\begin{split}
		2 \frac{\partial \phi}{\partial \delta q} &= \frac{1}{C \delta\alpha} - \frac{1}{C^2 \delta\alpha^2} + \left( \alpha_c - \delta \alpha \right) \left[ \frac{H\left(-\kappa \right)}{C \delta\alpha} + \frac{B(\kappa)}{C^2 \delta\alpha^2} \right] \\
		&= \frac{1}{C \delta\alpha} \left[ 1 + \alpha_c H\left( -\kappa \right) - \frac{B(\kappa)}{C}\right] + \frac{1}{C^2 \delta\alpha^2} \left[\alpha_c B(\kappa) - 1 \right] = 0 \,.
	\end{split}
\end{equation}
The first term gives the scaling of $\delta q$, the second gives us the critical capacity in terms of the margin~\cite{gardner1988The,gardner1988optimal}. 
\begin{equation}
	\label{eq::alphac_spherical}
	\alpha_c(\kappa) = \frac{1}{B(\kappa)} = \frac{1}{\kappa G\left( \kappa \right) + \left( \kappa^2 + 1 \right) H\left( -\kappa \right)}
\end{equation}
Notice that $\alpha_c = \frac{1}{B(\kappa)}$ is equivalent to imposing that the divergence $1/\delta q$ in the free entropy~\eqref{eq::phi} is eliminated at the critical capacity (so that it correctly goes to $-\infty$ in that limit). In particular for $\kappa = 0$ we get
\begin{equation}
	\alpha_c(\kappa = 0) = 2 \,,
\end{equation}
a results that has been derived rigorously by Cover in 1965~\cite{Cover1965}. In the right panel of Fig.~\ref{Fig::spherical_entropy_and_alphac} we plot $\alpha_c$ as a function of $\kappa$ obtained by~\eqref{eq::alphac_spherical}. 
It is important to mention that, since in the case $\kappa \le 0$ the model becomes non-convex, the RS estimate of the critical capacity~\eqref{eq::alphac_spherical} gives uncorrect results, even if it is un upper bound to the true value. We refer to~\cite{montanari2021tractability} for a rigorous upper bound to the critical capacity and to~\cite{baldassi2023typical} for the evaluation of $\alpha_c(\kappa)$ based on a 1RSB ansatz. However the correct result should be obtained by performing a full-RSB ansatz~\cite{Parisi1980,franz2017}.

\section{Landscape geometry}

One of the most important open puzzles in deep learning is to understand how the error and the loss landscape look like~\cite{epfl_workshop} especially as a function of the number of parameters, and how the shape of the learning landscape impacts the learning dynamics. So far there has been growing empirical evidence that, especially in the \emph{overparameterized regime}, where the number of parameters is much larger that the size of the training set, the landscape presents a region at low values of the loss with a large number of flat directions. For example, studies of the spectrum of the Hessian~\cite{sagun2016eigenvalues,sagun2017empirical} on a local minima or a saddles found by Stochastic Gradient Descent (SGD) optimization, show the presence of a large number of zero and near to zero eigenvalues. This region seem to be attractive for the gradient-based algorithms: the authors of~\cite{draxler2018} show that different runs of SGD end in the same basin by explicitly finding a path connecting them, in~\cite{entezari2022,pittorino22} it was shown that very likely even the whole straight path between two different solutions lies at zero training error, i.e. they are \emph{linear mode connected}. Also, an empirical evidence of an implicit bias of SGD towards flat regions have been shown in~\cite{FengYuhai}. Study of the limiting dynamics of SGD~\cite{Chen2022,kunin2021rethinking} unveil the presence of a diffusive behaviour in weight space, once the training set has been fitted. Simple 2D visualization of the loss landscape~\cite{LiVisualizing2018}, provided insights into how commonly employed machine learning techniques for improving generalization also tend also to smooth the landscape. One of the most recent algorithms, Sharpness Aware Minimization (SAM)~\cite{foret2021sharpnessaware}, explicitly designed to target flat regions within the loss landscape, consistently demonstrates improved generalization across a broad spectrum of datasets. In~\cite{baity2019comparing} it was suggested numerically that moving from the over to the underparameterized regime, gradient based dynamics suddendly becomes glassy. This observation raises the intriguing possibility of a phase transition occurring between these two regimes.

In this section we review some simple results obtained on one layer models concerning the characterization of the flatness of different classes of solutions. We will consider, the paradigmatic case of the binary perceptron, but, as we will see, the definitions of the quantities are general, and can be also used in the case of continuous weights. We will see that in the overparameterized regime (i.e. $N\gg P$) solutions located in a  very wide and flat region exist. However, as the number of constraints is increased, this region starts to shrink and at a certain point it breaks down in multiple pieces. This, in binary models, is responsible of algorithmic hardness and glassy dynamics.


\subsection{Local Entropy}
\begin{figure}[t]
	\begin{centering}
		\includegraphics[width=0.5\columnwidth]{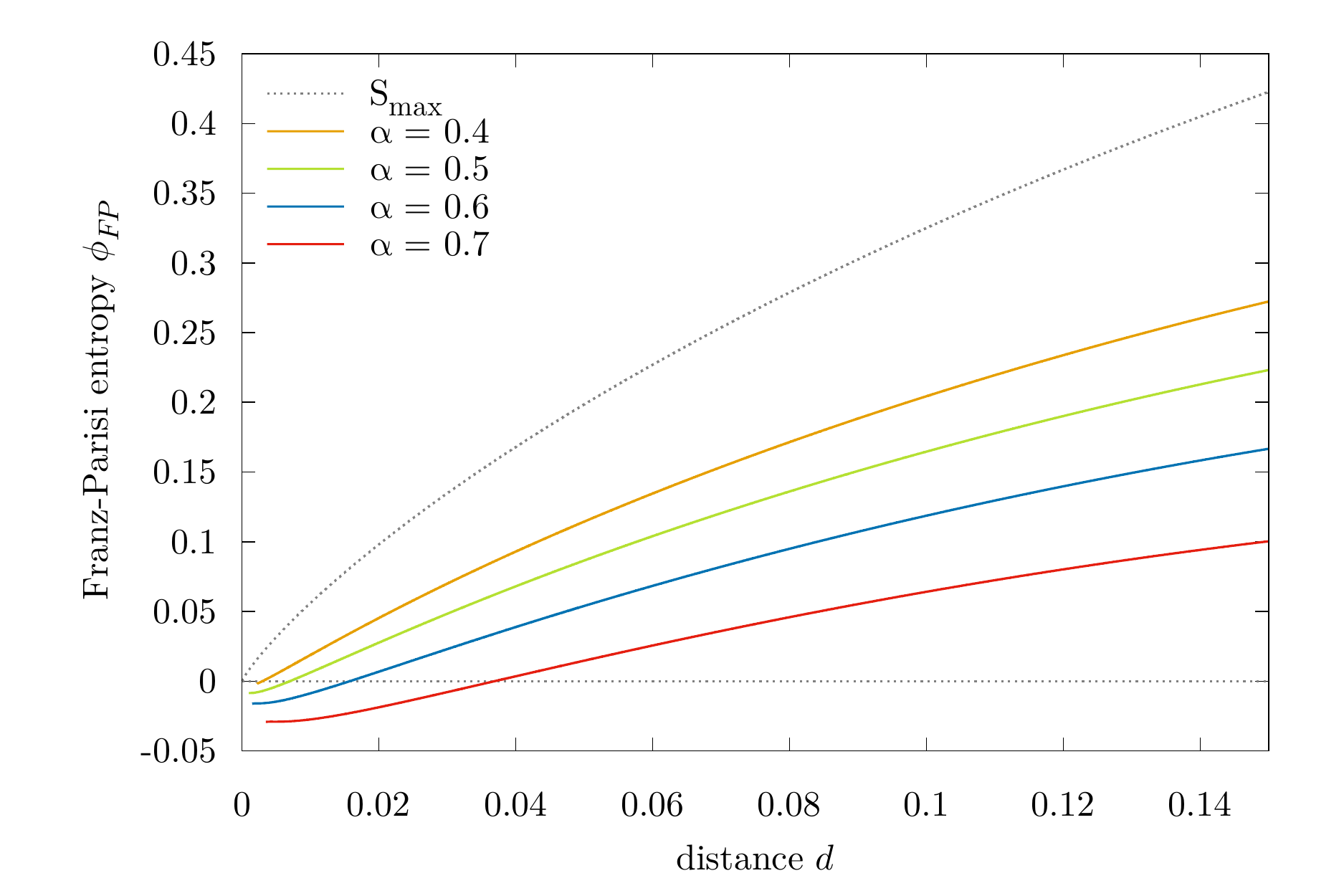}
		\includegraphics[width=0.5\columnwidth]{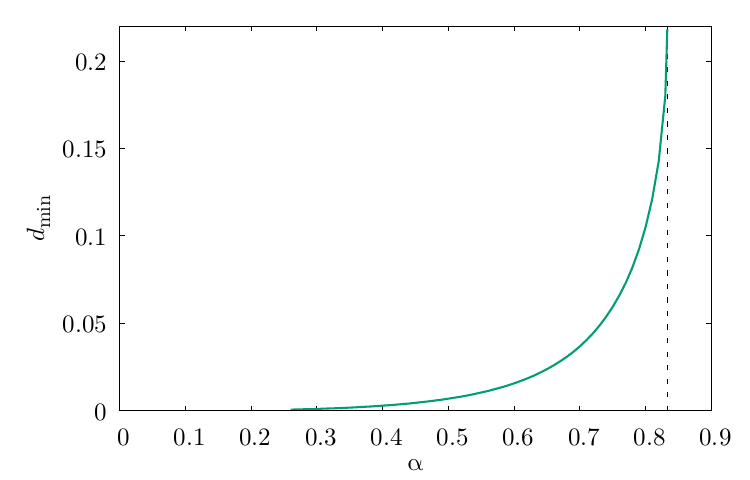}
	\end{centering}
	\caption{\textbf{Binary perceptron}. {\bf{Left}}: Averaged local entropy of typical solutions as a function of distance, for several values of $\alpha$. {\bf{Right}}: Distance $d_{\text{min}}$ for which the Franz Parisi entropy $\phi_{\text{FP}}$ is zero as a function of $\alpha$. At the SAT/UNSAT transition (dashed vertical line) the minimal distance to the closest solutions coincides with the typical distance between solution $d \simeq 0.218$.}
	\label{Fig::Huang_Kabashima} 
\end{figure}

In order to quantify the flatness of a given solutions, several measures can be employed. One such obvious measure is the spectrum of the Hessian; however this quantity is not trivial to study analytically. Here we will employ the so called \emph{local entropy}~\cite{baldassi2015subdominant} measure. Given a configuration $\tilde{\boldsymbol{w}}$ (in general it may not be a solution), its local entropy is defined as
\begin{equation}
	\mathcal{S}_{\mathcal{D}}(\tilde{\boldsymbol{w}}, d; \kappa) = \frac{1}{N} \ln \mathcal{N}_{\mathcal{D}}(\tilde{\boldsymbol{w}}, d; \kappa)
\end{equation}
where $\mathcal{N}_{\mathcal{D}}(\tilde{\boldsymbol{w}}, d; \kappa)$ is a \emph{local Gardner volume}
\begin{equation}
	\label{eq::local_Garnder_volume}
	\mathcal{N}_{\mathcal{D}}(\tilde{\boldsymbol{w}}, d; \kappa) \equiv \int d \mu(\boldsymbol{w}) \, \mathbb{X}_{\mathcal{D}}(\boldsymbol{w}; \kappa) \, \delta( N(1-2d) - \boldsymbol{w} \cdot \tilde{\boldsymbol{w}})
\end{equation}
which counts\footnote{In the spherical case it measures a volume.} solutions having margin at least $\kappa$ at a given distance $d$ from $\tilde{\boldsymbol{w}}$. In the binary perceptron, we will set for simplicity $\kappa = 0$ in the above quantity, since for $\kappa=0$ the problem is already non-convex. 
In~\eqref{eq::local_Garnder_volume} we have used the relation between overlap and distance of~\eqref{eq::distance_overlap_relation} to impose an hard constraint between $\tilde{\boldsymbol{w}}$ and $\boldsymbol{w}$. For any distance, the local entropy is bounded from above by the \emph{total} number of \emph{configurations} at distance $d$. This value, that we will call $\mathcal{S}_{\text{max}}$, is attained at $\alpha = 0$, i.e. when we do not impose any constraints. Moreover, since every point on the sphere is equivalent, $\mathcal{S}_{\text{max}}$ does not depend on $\tilde{\boldsymbol{w}}$ and $\kappa$ and it reads\footnote{In the spherical case an equivalent analytical formula can be derived, see~\cite{baldassi2023typical}.}
\begin{equation}
	\label{eq::Smax}
	\mathcal{S}_{\text{max}}(d) = -d \ln d - (1-d) \ln (1-d) 
\end{equation}
which is of course always non-negative.
In full generality, we are interested in evaluating the local entropy of solutions $\tilde{\boldsymbol{w}}$ that have margin $\tilde{\kappa}$ and that are sampled from a probability distribution $P_{\mathcal{D}}(\tilde{\boldsymbol{w}}; \tilde{\kappa})$. We are interested in computing the typical local entropy of those class of solutions, that is obtained averaging $\mathcal{S}_{\mathcal{D}}$ over $P_{\mathcal{D}}$ and over the dataset, i.e.
\begin{equation}
	\label{eq::FP_entropy}
	\phi_{\text{FP}}(d; \tilde{\kappa}, \kappa) = \left\langle \int d\mu(\tilde{\boldsymbol{w}}) \, P_{\mathcal{D}}(\tilde{\boldsymbol{w}}; \tilde{\kappa}) \, \mathcal{S}_{\mathcal{D}}(\tilde{\boldsymbol{w}}, d; \kappa) \right\rangle_{\mathcal{D}}
\end{equation}
This ``averaged local entropy'' is usually called \emph{Franz-Parisi entropy} in the context of mean field spin glasses~\cite{franz1995recipes}. In the following we will consider $P_{\mathcal{D}}$ as the flat measure over solutions with margin $\tilde{\kappa}$
\begin{equation}
	\label{eq::flat_measure_kt}
	P_{\mathcal{D}}(\tilde{\boldsymbol{w}}; \tilde{\kappa}) = \frac{\mathbb{X}_{\mathcal{D}}(\tilde{\boldsymbol{w}}, \tilde{\kappa})}{\int d\mu(\tilde{\boldsymbol{w}}) \, \mathbb{X}_{\mathcal{D}}(\tilde{\boldsymbol{w}}, \tilde{\kappa})}
\end{equation}
The first analytical computation of~\eqref{eq::FP_entropy} was performed by Huang and Kabashima~\cite{huang2014origin} in the binary perceptron by using the replica method (using steps similar to the ones done in Section~\ref{sec::Replica_Method}, even if thet are much more involved). They considered the case of typical solutions, i.e. $\tilde{\kappa} = 0$. A plot of $\phi_{\text{FP}}$ as a function of distance is shown in the left panel of Fig.~\ref{Fig::Huang_Kabashima} for several values of $\alpha$. It exists a neighborhood of distances $d \in [0, d_{\text{min}}]$ around $\tilde{\boldsymbol{w}}$ for which the local entropy is negative, which is unphysical in the binary case. The authors of~\cite{huang2014origin} also showed analytically that $\phi_{\text{FP}}(d = 0) = 0$\footnote{This should be expected: if you are at zero distance from $\tilde{\boldsymbol{w}}$, the only solution to be counted is $\tilde{\boldsymbol{w}}$ itself!} and that for $d\to 0$
\begin{equation}
	\frac{\partial \phi_{\text{FP}}}{\partial d} = \alpha C d^{-1/2} + O(\ln d)
\end{equation}
where $C$ is a \emph{negative} constant. This tells us that $d_{\text{min}}>0$ \emph{for any} $\alpha>0$. 
This suggests that typical solutions are always \emph{isolated}, meaning that there is a value $d_{\text{min}}$ below which no solution can be found\footnote{In principle, the fact that the local entropy is negative suggest that only a \emph{subestensive} number of solutions can be found below $d_{\text{min}}$. Abbe and Sly~\cite{abbe2021proof} have also proved that in a slight variation of the model (the so called \emph{symmetric binary perceptron}), that actually \emph{no} solution can be found with probability one within a distance $d_{\text{min}}$. See also~\cite{perkins2021frozen}.}, no matter what the value of the constraint density is. Notice that, since the overlap is normalized by $N$ as in~\eqref{eq::overlap}, in order to go from $\tilde{\boldsymbol{w}}$ to the closest solution, one should flip an \emph{extensive} number of weights. The plot of $d_{\text{min}}$ is shown, as a function of $\alpha$, in the right panel of Fig.~\ref{Fig::Huang_Kabashima}.

The picture, however is far from being complete. If the landscape is composed only by those point-like solutions, intuition would suggest that finding solutions should be an hard optimization problem. 
Indeed, from the rigorous point of view, it has been shown in slight variations of the present model, such as the symmetric binary perceptron~\cite{Gamarnik2022,gamarnik2023geometric}, that algorithmic hardness is expected if the problem at hand verifies the so called \emph{Overlap Gap Property} (OGP)~\cite{gamarnik2021overlap,elAlaoui2021optimization}. Informally, an optimization problem possesses OGP if picking \emph{any} two solutions the overlap distribution between them exhibits a gap, i.e. they can be either close or far away from each other, but can't be in some interval in between. 

In the binary perceptron model that we have analyzed, however, very simple algorithms such as the ones based on message passing~\cite{Braunstein2006,baldassi2007efficient} or gradient-based methods~\cite{baldassi2022learning,baldassi2023typical} find solutions easily. This is not in constrast with OGP: it simply means that there should exists in the landscape other types of solutions, that are not isolated and that algorithms are able to find. We remark here that OGP is defined \emph{for any type} of solution one can sample; so far we have only focused on typical solutions.    In~\cite{baldassi2015subdominant} it was shown that non-isolated solutions indeed exists, but they are exponentially rarer (``subdominant''). 
In order to target those solutions one should give a larger statistical weight to those $\tilde{\boldsymbol{w}}$ that are surrounded by a larger number of solutions. This led~\cite{baldassi2015subdominant} to choose the measure
\begin{equation}
	\label{eq::P_largest_le}
	P_{\mathcal{D}}(\tilde{\boldsymbol{w}}; d) = \frac{e^{y N \mathcal{S}_{\mathcal{D}}(\tilde{\boldsymbol{w}}; d)}}{\int d\mu(\tilde{\boldsymbol{w}}) \, e^{y N\mathcal{S}_{\mathcal{D}}(\tilde{\boldsymbol{w}}; d)}} \,.
\end{equation}
where $y$ is a parameter (analogous to inverse temperature) that assigns larger statistical weight to solutions with high local entropy the larger it is. In the $y\to \infty$ limit, this measure focuses on solutions with the \emph{highest} local entropy for a given value of the distance $d$; notice, indeed, that this probability measure, depends explicitly on $d$, meaning that the particular type of solution $\tilde{\boldsymbol{w}}$ sampled changes depending on the value of $d$ chosen. In the same work it was shown not only that those high local entropy regions exist, but also that, in the teacher-student setting, they have better generalization properties with respect to typical, isolated solutions. For some rigorous results concerning the existence of those regions in similar binary perceptron models, see~\cite{abbe2021binary}. It was then shown that similar atypical clustered region play a similar algorithmic role in other optimization problems, such as coloring~\cite{Budzynski_2019,Cavaliere_2021}.  
Those results suggested the design of new algorithms based on message passing that explicitly exploit local entropy maximization in order to find very well-generalizing solutions~\cite{baldassi2016unreasoanable}. One of such algorithms is called focusing Belief-Propagation (fBP). 

\begin{figure}[t]
	\centering  
	\includegraphics[width=0.7\columnwidth]{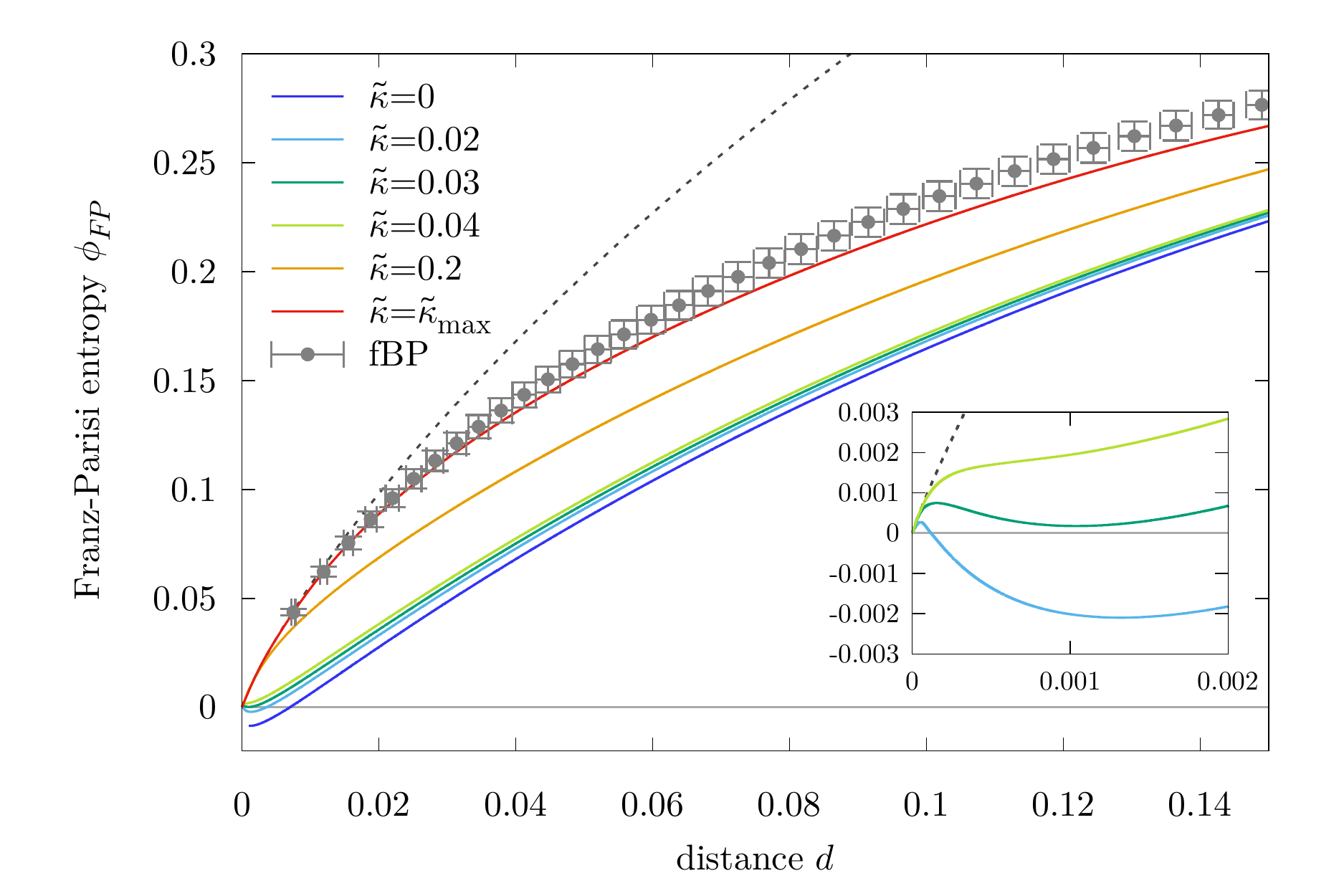}
	\caption{Averaged local entropy of solutions sampled with the flat measure over solutions with margin $\tilde{\kappa}$. The dashed line represents $\mathcal{S}_{\text{max}}(d)$ as in~\eqref{eq::Smax}. Notice how the local entropy changes from being non-monotonic to be monotonic as one keeps increasing $\tilde{\kappa}$. The grey points corresponds to the local entropy computed numerically from solutions found by the fBP algorithm, which it has been explicitly constructed to target the \emph{flattest} regions in the landscape. Especially at small distances, there is a good agreement with the local entropy of maximum margin solutions obtained from theory (red line). Reprinted from~\cite{baldassi2021unveiling}.}
	\label{Fig::FP_margin} 
\end{figure}

A simpler way of finding the high local entropy regions is by using~\eqref{eq::flat_measure_kt} 
with $\tilde{\kappa}$ stricly larger than zero~\cite{baldassi2021unveiling}.
Indeed, the property of being robust to small noise perturbation in the input is related to the flatness of the energy landscape in the neighborhood of the solution. 
This type of approach not only produces a phenomenology similar to~\cite{baldassi2015subdominant}, but it also helps to unravel the structure of high local entropy regions in neural networks, as we shall see.
Increasing the amount of robustness $\tilde{\kappa}$, we therefore intuitively expect to target larger local entropy solutions. As shown in Fig.~\ref{Fig::FP_margin}, this is indeed what one finds: as one imposes $\tilde \kappa > 0$, there always exists a neighborhood of $d=0$, with positive local entropy (i.e. those solutions are surrounded by an \emph{exponential} number of solutions). As shown in the inset of the same figure, the cluster is also very dense: for small $d$, the local entropy curve is indistinguishable from the total log-number of configurations at that distance $\mathcal{S}_{\text{max}}$. 
As one increases $\tilde{\kappa}$ from 0 one can see that one starts to sample different kind of regions in the solutions space.
Firstly, if $0 < \tilde{\kappa}<\tilde{\kappa}_{\text{min}}(\alpha)$ the local entropy is negative in an interval of distances $d \in [d_1 , d_2]$ with $d_1 > 0$: no solutions can be found in a spherical shell of radius $d \in [d_1 , d_2]$.
Secondly, if $\tilde{\kappa}_{\text{min}}(\alpha) <
\tilde{\kappa} < \tilde{\kappa}_u (\alpha)$ the local entropy is positive but non-monotonic. This means that typical solutions
with such $\tilde{\kappa}$ are immersed within small regions that
have a characteristic size: they can be considered as isolated (for $\tilde{\kappa}< \tilde{\kappa}_{\text{min}}$) or nearly isolated (for $\tilde{\kappa}> \tilde{\kappa}_{\text{min}}$) balls. Finally, for $\tilde{\kappa} > \tilde{\kappa}_u$, the local entropy is monotonic: this suggests that typical solutions with large enough $\tilde{\kappa}$ are immersed in dense regions that do not seem to have a characteristic size and may extend to very large scales. The local entropy curve having the highest local entropy at a given value of $\alpha$ is obtained by imposing the \emph{maximum} possible margin $\kappa_{\text{max}}(\alpha)$, i.e. the margin for which the entropy computed in section~\ref{sec::RS} vanishes\footnote{The $\kappa_{\text{max}}(\alpha)$ curve is the inverse of $\alpha_c(\kappa)$.}.

Therefore high margin solutions are not only rarer and closer to each other with respect to typical solutions (as examined in section~\ref{sec::Gardner_computation}), but tend to focus on regions surrounded by lower margin, which in turn are surrounded by many other solutions having even lower margin and so on and so forth. The flat regions in the landscape can be though to be formed by the coalescence of minima corresponding to high-margin classifications.

It is worth mentioning that in~\cite{baldassi2023typical} the same type of approach was applied to the simplest non-convex but continuous weights problem: the negative spherical perceptron. Although in the spherical case the most probable, typical solutions are not completely isolated\footnote{In the context of a spherical model, a solution would be isolated if the Franz-Parisi entropy goes to $-\infty$ for a $d_{\text{min}} >0$.}, a similar phenomenology is valid: higher margin solutions have always a larger local entropy. Evidence of the existence of these large local entropy regions has also been established in the context of the one hidden layer neural networks~\cite{relu_locent,baldassi2020shaping}. These studies also reveal that the use of the cross-entropy loss~\cite{baldassi2020shaping}, ReLU activations~\cite{relu_locent}, and even regularization of the weights~\cite{baldassi2020wide} influence the learning landscape by inducing wider and flatter minima.

\subsection{Algorithmic hardness}

\begin{figure}[t]
	\begin{centering}
		\includegraphics[width=0.493\columnwidth]{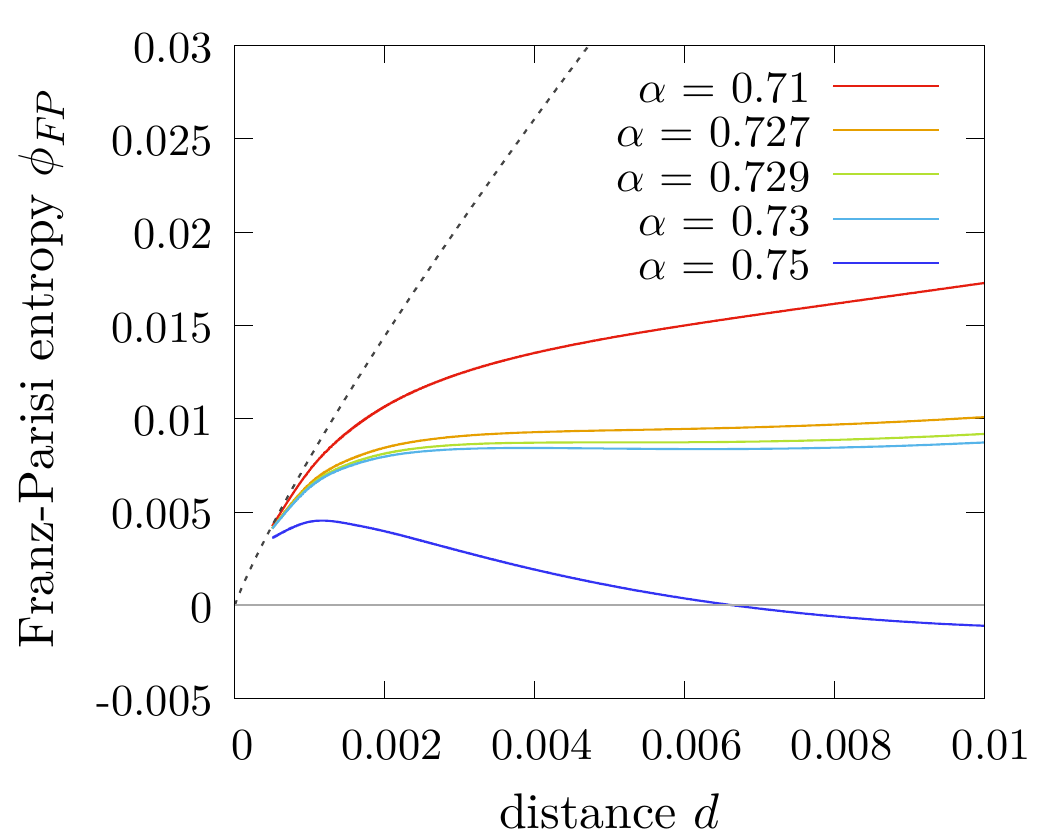}
		\includegraphics[width=0.493\columnwidth]{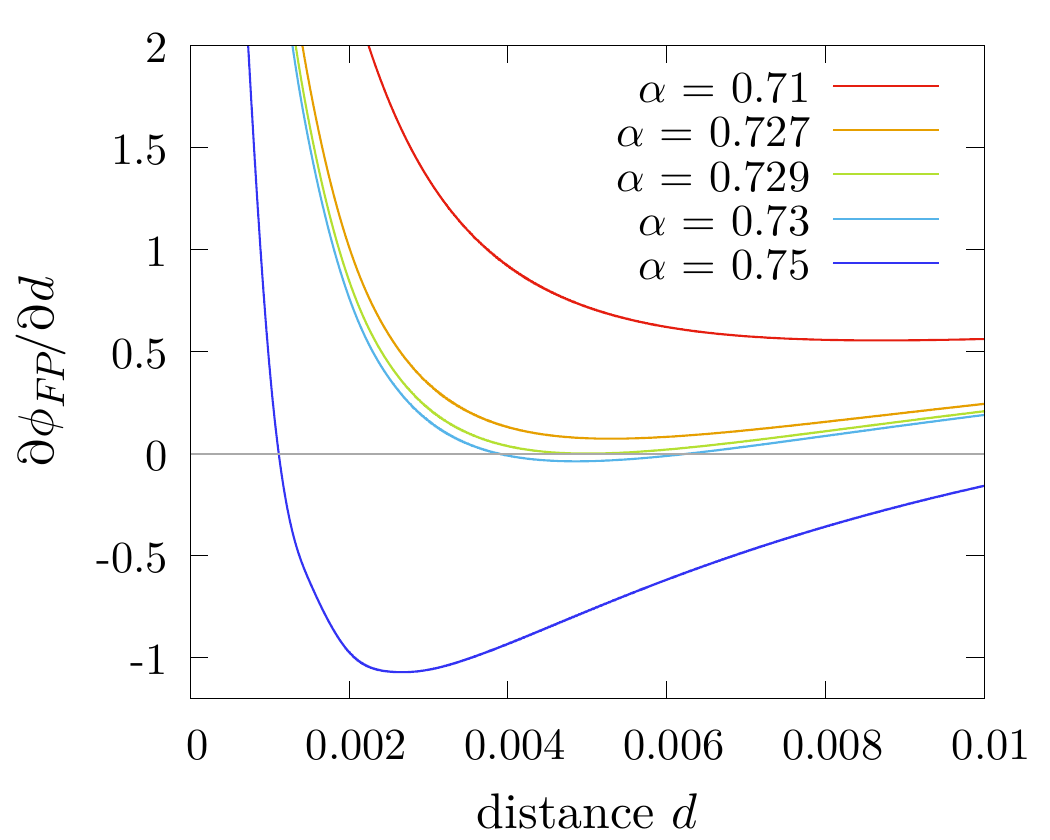}
	\end{centering}
	\caption{Averaged local entropy profiles of typical maximum margin solutions (left panel) and its derivative (right panel) as a function of the distance. The dashed line again represents $\mathcal{S}_{\text{max}}(d)$. Different values of $\alpha$ are displayed: for $\alpha=0.71$ and $0.727$ the entropy is monotonic, i.e. it has a unique maximum at large distances (not visible). For $\alpha = \alpha_{\text{LE}} \simeq 0.729$ the Franz-Parisi entropy becomes non-monotonic (its derivative with respect to the distance develops a new zero). The entropy becomes negative in a range of distances not containing the origin for $\alpha>\alpha_{\text{OGP}} \simeq 0.748$. Reprinted from~\cite{baldassi2021unveiling}.}
	\label{Fig::FP_kmax} 
\end{figure}

One can then explore how the highest local entropy curves evolve as a function of $\alpha$. The procedure works as follows. Firstly, for a given value of $\alpha$, we compute the maximum margin $\kappa_{\text{max}}$. Secondly, we plot the corresponding local entropy curve $\phi_{\text{FP}}$ as a function of $d$. Finally, we repeat the process using another value of $\alpha$. The outcome is plotted for the binary perceptron in the left panel of Fig.~\ref{Fig::FP_kmax}. 
As expected from the previous section, for low values of $\alpha$ the Franz-Parisi entropy is monotonic. However, as we keep increasing $\alpha$ the curve becomes non-monotonic: as shown in the right panel of Fig.~\ref{Fig::FP_kmax} the derivative of $\phi_{\text{FP}}$ with respect to $d$ develops a zero at small distances. This critical value of the constrained density has been called ``local entropy'' transition $\alpha_{\text{LE}}$, and it separates a phase where we can find a solution $\tilde{\boldsymbol{w}}$ that is located inside a region that extends to very large distance $\alpha < \alpha_{\text{LE}}$, from one $\alpha > \alpha_{\text{LE}}$ where it can't be found. Above another critical value $\alpha_{\text{OGP}}$ of the constrained density, only the ``completely isolated ball'' phase exists: all the high-margin solutions, even if they remain surrounded by an exponential number of lower margin solutions up to the SAT/UNSAT transition, are completely isolated between each other.

The local entropy transition has been shown in the binary perceptron~\cite{baldassi2015subdominant, baldassi2021unveiling} to be connected with the onset of~\emph{algorithmic hardness}: no algorithm has currently been found to be able to reach zero training error for $\alpha > \alpha_{\text{LE}}$. Surprisingly, $\alpha_{\text{LE}}$, which has been derived as a threshold marking a profound change in the geometry of regions of higher local entropy, also acquires the role of an insurmountable \emph{algorithmic} barrier. Similar algorithmic thresholds have been found to exists in other binary neural network models~\cite{baldassi2022learning}. 
Notice that, OGP is expected to hold for $\alpha > \alpha_{\text{OGP}}$; indeed if the Franz-Parisi entropy displays a gap for the $\kappa_{\text{max}}$ curve, it will also exhibit an even larger one for every $\tilde{\kappa} \in [0, \kappa_{\text{max}})$\footnote{To be precise the correct value of $\alpha_{\text{OGP}}$ (and of $\alpha_{\text{LE}}$), can be obtained by sampling the reference with the \emph{largest} local entropy at any distance, i.e. by using~\eqref{eq::P_largest_le}.}. In the binary perceptron model, $\alpha_{\text{LE}}$ and $\alpha_{\text{OGP}}$ are very close, so it is really difficult to understand which of the two prevents algorithms to find solutions in the infinite size limit.

In spherical non-convex models, such as the negative margin perceptron, even if the local entropy transitions can be identified similarly, it has been shown to be not predictive of the behaviour algorithms, but rather, there is numerical evidence showing that it demarks a region where the solution space is connected from one where it is not~\cite{baldassi2023typical}. In the same work, evidence has been given that smart algorithms are able to reach the SAT/UNSAT transition. In~\cite{elAlaoui2022algorithmic} the authors, interestingly, developed an algorithm that is proved to be able to reach the SAT/UNSAT transition, provided that the OGP does not hold. A proof of the lack of OGP in the spherical negative perceptron, however, is still lacking.

\section{Linear mode connectivity}

Another important line of research that has recently emerged in machine learning is the characterization of the connectivity between different solutions, i.e. the existence of a path lying a zero training error that connects them. Numerous studies~\cite{goodfellow2014qualitatively,draxler2018,frankle2020revisiting,vlaar2022can,wang2023plateau,garipov2018} have been started analyzing the so called \emph{linear mode connectivity}, i.e. the particular case of a simple straight path. 

The first statistical mechanics study of the behavior of the training error on the \emph{geodesic path} connecting two solutions has been done in~\cite{annesi2023star} for the negative spherical perceptron problem. Interestingly, a complete characterization of the \emph{shape} of the solution space of the spherical negative perceptron has been produced. It is the aim of this section to briefly explain the analytical technique and the main results of the work.  
\begin{figure}[t]
	\begin{centering}
		\includegraphics[width=0.493\columnwidth]{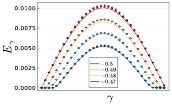}
		\includegraphics[width=0.493\columnwidth]{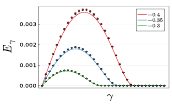}
	\end{centering}
	\caption{Average barrier on the geodesic path connecting two solutions $\boldsymbol{w}_1$ and $\boldsymbol{w}_2$ sampled with margin $\kappa_1$, $\kappa_2 \ge \kappa$. In both plots $\kappa = -0.5$ and $\alpha=1$. Lines are the theoretical predictions; the points corresponds to numerical simulations performed by finding $\boldsymbol{w}_1$ and $\boldsymbol{w}_2$ using Simulated Annealing and extrapolating $E_\gamma$ to large sizes. {\bf Left panel}: $\kappa_1 = \kappa_2 = \tilde{\kappa} \ge \kappa$ case. {\bf Right panel}: case $\kappa_1 = \kappa$, and $\kappa_2 = \tilde{\kappa} > \kappa$. Reprinted from~\cite{annesi2023star}}
	\label{Fig::barriers} 
\end{figure}

Suppose that we are given a value of $\kappa < 0$ and we need to satisfy the constraints~\eqref{eq::margin} using spherical weights. We sample two solutions $\boldsymbol{w}_1$, $\boldsymbol{w}_2$ from the probability distribution~\eqref{eq::P_largest_le} using as margin respectively $\kappa_1, \, \kappa_2 \ge \kappa$. Since the model is defined on the sphere, the straight path between $\boldsymbol{w}_1$ and $\boldsymbol{w}_2$ lies out of the sphere itself; therefore we project it on the sphere, obtaining a set of weights $\boldsymbol{w}_\gamma$ that lie on the minimum length (i.e. the geodesic) path joining $\boldsymbol{w}_1$ and $\boldsymbol{w}_2$ Then we can define the \emph{geodesic path} between $\boldsymbol{w}_1$ and $\boldsymbol{w}_1$: 
\begin{equation}
	\boldsymbol{w}_{\gamma} = \frac{\sqrt{N} \left(\gamma \boldsymbol{w}_1 + (1-\gamma) \boldsymbol{w}_2\right)}{\lVert \gamma \boldsymbol{w}_1 + (1-\gamma) \boldsymbol{w}_2 \rVert} \,, \qquad \gamma \in [0, 1]
\end{equation}
Finally we compute the average training error of $\boldsymbol{w}_\gamma$ i.e. the fraction of errors on the training set, averaged over the sampled $\boldsymbol{w}_1$, $\boldsymbol{w}_2$ and over the realization of the dataset
\begin{equation}
	\label{eq::general_energybarrier}
	E_{\gamma} = \lim_{N\to+\infty} \,\mathbb{E}_{\mathcal{D}}\,\left\langle \frac{1}{P} \sum_{\mu=1}^P \Theta\left( - \boldsymbol{w}_{\gamma} \cdot \boldsymbol{\xi}^\mu + \kappa \sqrt{N} \right)\right\rangle_{\kappa_1, \kappa_2}.
\end{equation}
In the previous expression the average $\left\langle \bullet \right \rangle_{\kappa_1, \kappa_2}$ is over the product of the two Gibbs ensembles~\eqref{eq::P_largest_le} from which $\boldsymbol{w}_1$ and $\boldsymbol{w}_2$ are sampled from. $E_\gamma$ can be computed by using replica method. 
Depending on the values of the margin $\kappa_1$, $\kappa_2$ of $\boldsymbol{w}_1$ and $\boldsymbol{w}_2$ we can sample different regions of the solution space for a fixed value of $\alpha$. We mainly have three cases:
\begin{itemize}
	\item $\kappa_1 = \kappa_2 = \kappa$: the two solutions are typical. In this case, for every $\gamma\in[0,1]$, $E_\gamma >0$, see red line in the left panel of Fig.~\ref{Fig::barriers}. 
	\item $\kappa_1 = \kappa_2 = \tilde{\kappa} > \kappa$: the two solutions are atypical and have the same margin. As can be seen in the left panel of Fig.~\ref{Fig::barriers}, if $\tilde{\kappa}$ is slightly above $\kappa$, the maximum energy barrier is still non-zero, but in the neighborhood of $\boldsymbol{w}_1$ and $\boldsymbol{w}_2$ a region at zero training error appears. The size of this region increases with $\tilde{\kappa}$. If $\tilde{\kappa} > \kappa_\star(\alpha)$ the maximum barrier is zero, i.e. $\boldsymbol{w}_1$ and $\boldsymbol{w}_2$ are \emph{linear mode connected}. 
	\item $\kappa_1 = \kappa$ and $\kappa_2 = \tilde{\kappa} > \kappa$: one solution is typical and the other is atypical and $E_\gamma$ is asymmetric with respect to $\gamma = 1/2$. $E_\gamma$ is shown in the right panel of Fig.~\ref{Fig::barriers}, for several values of $\tilde{\kappa}$. As one keeps increasing $\tilde{\boldsymbol{w}}$, the maximum of the barrier decreases, and for $\tilde{\kappa} > \kappa_{\text{krn}}(\alpha) > \kappa_{\star}(\alpha)$ the two solutions become linear mode connected. It can be also shown analytically that if $\tilde{\kappa}>\kappa_{\text{krn}}(\alpha)$ than $\boldsymbol{w}_2$ is linear mode connected to another solution having margin $\kappa_1>\kappa$.
\end{itemize}
The picture exposed above holds in the overparameterized regime. 
In particular, the result obtained in the last point tells us that the solution space $\mathcal{A}$ of the spherical negative perceptron is \emph{star-shaped}, since there exists a subset $\mathcal{C} \subset \mathcal{A}$, such that for any $\boldsymbol{w}_{\star} \in \mathcal{C}$ then geodesic path from $\boldsymbol{w}_{\star}$ to \emph{any} other solution $\boldsymbol{w}$ lies entirely in $\mathcal{A}$, i.e.
\begin{equation}
\left\{ \gamma \boldsymbol{w} + (1-\gamma)\boldsymbol{w}_{\star}; \; \, \gamma \in [0,1]\right\} \subset \mathcal{A} \,.
\end{equation} 
\begin{figure}[t]
	\begin{centering}
		\includegraphics[width=0.3\columnwidth]{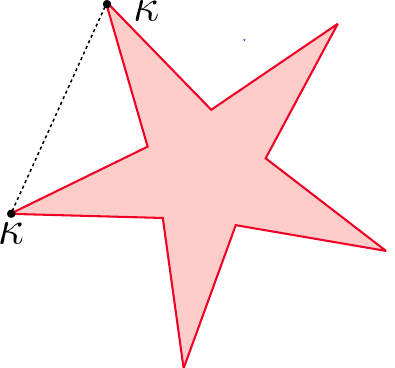}
		\hfill
		\includegraphics[width=0.3\columnwidth]{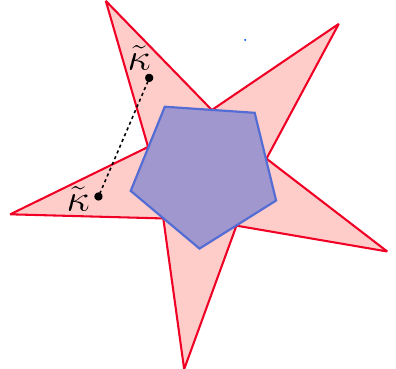}
		\hfill
		\includegraphics[width=0.3\columnwidth]{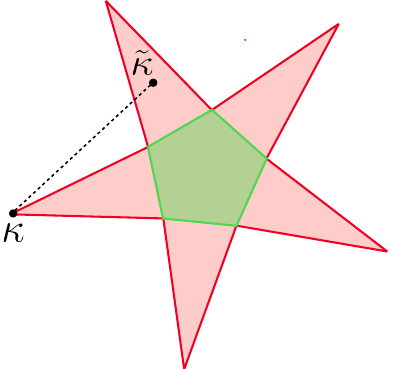}
	\end{centering}
	\caption{A two-dimensional view of the star-shaped manifold of solutions of the spherical negative perceptron in the overparameterized regime. When one samples typical solutions (left panel) the training error along geodesic path (dashed black line) is strictly positive. Increasing the margin of both solutions they get closer (see left panel of Fig.~\ref{Fig::distance}), and on geodesic path it appears a region at zero training error in their neighborhood. If the two solutions have margin larger than $\kappa_{\star}$, the maximal barrier goes to zero, and the two solutions are located in the blue area. If the margin of one solution $\boldsymbol{w}_{\star}$ is larger than $\kappa_{\text{krn}}> \kappa_{\star}$, then it will be located in the \emph{kernel} region (green area, right panel); from there it can ``view'' any other solution, since the geodesic path is all at zero training error. }
	\label{Fig::star} 
\end{figure}
The subset $\mathcal{C}$ is called \emph{kernel} of the star-shaped manifold. In Fig.~\ref{Fig::star} we show an intuitive, 2D picture of the space of solutions, showing a schematic interpretation of the results exposed. If we sample typical solutions we are basically sampling the tips of the star (see left panel), so that the geodesic path connecting them lies entirely at positive training error. Remind that those solutions are the largest in numbers and are located at a larger typical distance with respect to higher margin ones. As one increases the margin of $\boldsymbol{w}_1$, $\boldsymbol{w}_2$ they come closer together, following the arms of the star to which they belong to. If the two margin are large enough, i.e. $\kappa_1$, $\kappa_2 > \kappa_\star(\alpha)$ they are linear mode connected, and they lie in the blue region of middle panel.
If the margin of one of the two solutions is even larger, $\kappa_1 > \kappa_{\text{krn}}(\alpha)$, then it will be located in the kernel, which is depicted in green in the right panel of Fig.~\ref{Fig::star}. We refer to~\cite{annesi2023star} to a more in-depth discussion of the attractiveness of the kernel region to gradient-based algorithms.

\bibliography{references.bib}


\end{document}